\UseRawInputEncoding
\documentclass[conference, compsoc]{IEEEtran}
\IEEEoverridecommandlockouts
\usepackage[backref]{hyperref}
\usepackage{cite}
\usepackage{amsmath, amssymb, amsfonts}
\usepackage{algorithmic}
\usepackage[dvipdfmx]{graphicx}
\usepackage{bmpsize}
\usepackage{textcomp}
\usepackage{xcolor}
\usepackage{wrapfig}
\usepackage{multirow}
\usepackage{multicol}
\def\BibTeX{{\rm B\kern-.05em{\sc i\kern-.025em b}\kern-.08em
    T\kern-.1667em\lower.7ex\hbox{E}\kern-.125emX}}

\begin{document}
\title{Enhancing Generalized Fetal Brain MRI Segmentation using A Cascade Network \\ with Depth-wise Separable Convolution and Attention Mechanism}

\author{\IEEEauthorblockN{Zhigao Cai$^1$ \quad Xing-Ming Zhao$^{1,2,3,}$\IEEEauthorrefmark{2}\thanks{\IEEEauthorrefmark{2} Corresponding author.}}
\IEEEauthorblockA{$^1$ Institute of Science and Technology for Brain-Inspired Intelligence, Fudan University, China\\
zgcai21@m.fudan.edu.cn, xmzhao@fudan.edu.cn \\
$^2$MOE Key Laboratory of Computational Neuroscience and Brain-Inspired Intelligence\\
$^3$MOE Frontiers Center for Brain Science, Shanghai, China\\
}}

\maketitle

\begin{abstract}
Automatic segmentation of the fetal brain is still challenging due to the health state of fetal development, motion artifacts, and variability across gestational ages, since existing methods rely on high-quality datasets of healthy fetuses. In this work, we propose a novel cascade network called CasUNext to enhance the accuracy and generalization of fetal brain MRI segmentation. CasUNext incorporates depth-wise separable convolution, attention mechanisms, and a two-step cascade architecture for efficient high-precision segmentation. The first network localizes the fetal brain region, while the second network focuses on detailed segmentation. We evaluate CasUNext on 150 fetal MRI scans between 20 to 36 weeks from two scanners made by Philips and Siemens including axial, coronal, and sagittal views, and also validated on a dataset of 50 abnormal fetuses. Results demonstrate that CasUNext achieves improved segmentation performance compared to U-Nets and other state-of-the-art approaches. It obtains an average Dice coefficient of 96.1\% and mean intersection over union of 95.9\% across diverse scenarios. CasUNext shows promising capabilities for handling the challenges of multi-view fetal MRI and abnormal cases, which could facilitate various quantitative analyses and apply to multi-site data.
\end{abstract}

\IEEEpeerreviewmaketitle


\section{Introduction}
Fetal Magnetic Resonance Imaging (MRI) plays a vital role in quantitative brain volumetry studies for prenatal diagnosis and assessing early human brain development\cite{1}. Accurate and automated segmentation of the fetal brain is crucial for various brain analysis tasks, including high-resolution reconstruction and cortical surface analysis. However, fetal MRI are often affected by severe and arbitrary motion artifacts, and clinical datasets frequently consist of abnormal developing brains. As a result, there is a lack of general methods for fetal brain segmentation, and existing approaches rely on diverse segmentation techniques that require extensive manual refinement, leading to time-consuming processes.\par
Methods relying on atlases have been wildly used in medical segmentation, but they may not generalize well to rapidly developing fetal brains and abnormal cases which have high anatomical variability outside the normal range\cite{5}.Over the years, with the successful application of neural networks in medical image processing and analysis\cite{9,10,11,12,13}, various strategies have emerged for automated segmentation and brain extraction from fetal T2-weighted MRI\cite{16,37,39}. However, existing approaches still face challenges in robustly handling the diverse scenarios. For instance, convolutional neural networks (CNNs) have been extensively employed for fetal brain extraction in conjunction with conditional random fields (CRFs) iteratively trained until convergence\cite{14}, whereas the model not designed for fetal MRI specifically may underperform. Salehi et al.\cite{16}proposed and evaluated a deep fully CNN based on a 2D U-Net, while single CNN-based model may overfit and not robustly handle multi-view scans (axial, coronal, sagittal) or motion artifacts outside the training distribution. Lou et al.\cite{17} presented an automatic brain image segmentation method utilizing a multi-stage 2D U-Net with deep supervision, operating slice-by-slice do not capitalize on inter-slice 3D relationships to refine boundaries or enforce volumetric consistency, especially with sparse noisy fetal MRI. Li et al.\cite{18} developed a two-step framework involving two fully convolutional networks (FCN) and an additional deep multiscale FCN for the extraction of fetal brains, while the model trained on a small dataset of 88 samples are prone to overfitting and may not generalize well to multi-institutional data or abnormal anatomy.\par
Advanced segmentation techniques were employed to obtain the final segmentation results. Ebner et al.\cite{19} introduced a four-stage fully automated fetal brain reconstruction framework employing CNNs for coarse-to-fine segmentation, requiring extensive manual intervention, refinement or parameter tuning, which is time-consuming for large clinical applications. Furthermore, Fusion Network\cite{20} and Confidence-Aware Network\cite{21} were proposed and utilized for post-processing of segmentation outcomes, but not focused on abnormal or pathological cases which may fail for atypical fetal brain. Overall, while these methods have achieved noteworthy results, they may not fully address the challenges specific to multi-site data and abnormal cases in fetal brain segmentation.\par
In this work, we aim to enhance the accuracy of fetal brain MRI segmentation by constructing a novel cascade framework based on ConvNext\cite{29} and Transformer\cite{25,26,27,28}. The proposed network, CasUNext, combines depth-wise separable convolution and inverted bottleneck with block ratio modification. This improvement not only enhances computational efficiency, but also enables a more compact model size. Furthermore, we incorporate an attention mechanism to optimize U-Net based backbone, facilitating the integration of low-level and high-level features. Our proposed method addresses the challenges associated with generalized processing multi-site data of different scales and abnormal fetal brains as well.\par
This paper is organized as follows: Section 2 presents a review of related work in the field of fetal brain MRI segmentation. In Section 3, we discuss the methodology and provide a detailed explanation of the proposed approach. The experimental setup, including the datasets and evaluation metrics, is outlined in Section 4, while Section 4 also presents the results and analysis of our experiments. Finally, Section 5 provides concluding remarks and discussions on the implications of our findings.\par


\section{Related Work}
Automated segmentation of fetal brain MRI has gained increasing research attention with the advancement of deep learning techniques, initial approaches focused on adapting standard architectures like U-Net\cite{15} and ResNet\cite{22}. Salehi et al.\cite{16} proposed a U-Net model augmented with auto-context features. Lou et al.\cite{17} presented a multi-stage U-Net pipeline with deep supervision for refinement. By applying residual connections, U-Net has achieved performance optimization by in medical image segmentation\cite{24}. More advanced network architectures have also emerged such as the attention-equipped U-Net by Ozan et al.\cite{36} which selectively focuses on relevant features. The cascaded segmentation paradigm has shown utility by separating localization and fine-grained segmentation steps\cite{19,20}. However, most methods evaluate on limited datasets often lacking diversity in gestational age, view angles, and pathologies. \par
Besides, ConvNext\cite{29} is a notable advancement based on the ResNet architecture enhancing the performance of CNN, and has achieved excellent results in tasks such as glioma\cite{30} and esophageal tumor segmentation\cite{24}. Then ConvUnext\cite{31} builds upon the success of ConvNext and applies to the U-Net architecture, leading to enhance image feature extraction and fusion capabilities. The aforementioned techniques have made significant contributions to medical image segmentation\par
Our work aims to improve generalization to diverse fetal scans by designing a cascade architecture, separable convolutions, and attention modules. We demonstrate state-of-the-art performance across multi-site data and abnormal cases.\par

\section{Method}
\subsection{Cascade network architecture}
The proposed CasUNext framework consists of two key components - a global localization network (Loc-Net) and a fine segmentation network (Seg-Net). Both networks share a similar overall architecture based on a convolutional neural network with an encoder-decoder structure. This cascaded design aims to improve the overall segmentation accuracy and empower the entire network to effectively handle complex image data.\par
\begin{figure}[ht]
  \centering
  \includegraphics[width=0.45\textwidth]{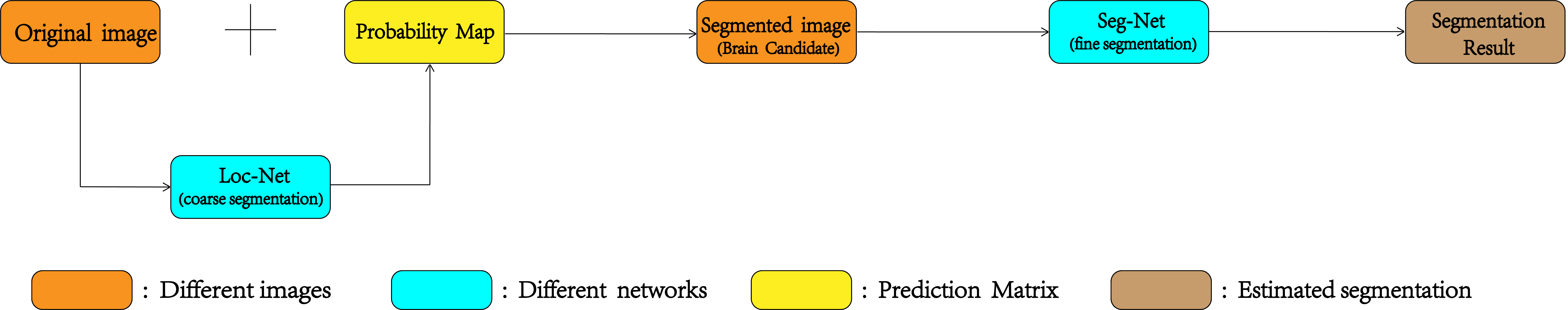}
  \caption{The algorithm flow of CasUnext}
  \label{fig:CasNet}
\end{figure}
\begin{figure}[ht]
  \centering
  \includegraphics[width=0.45\textwidth]{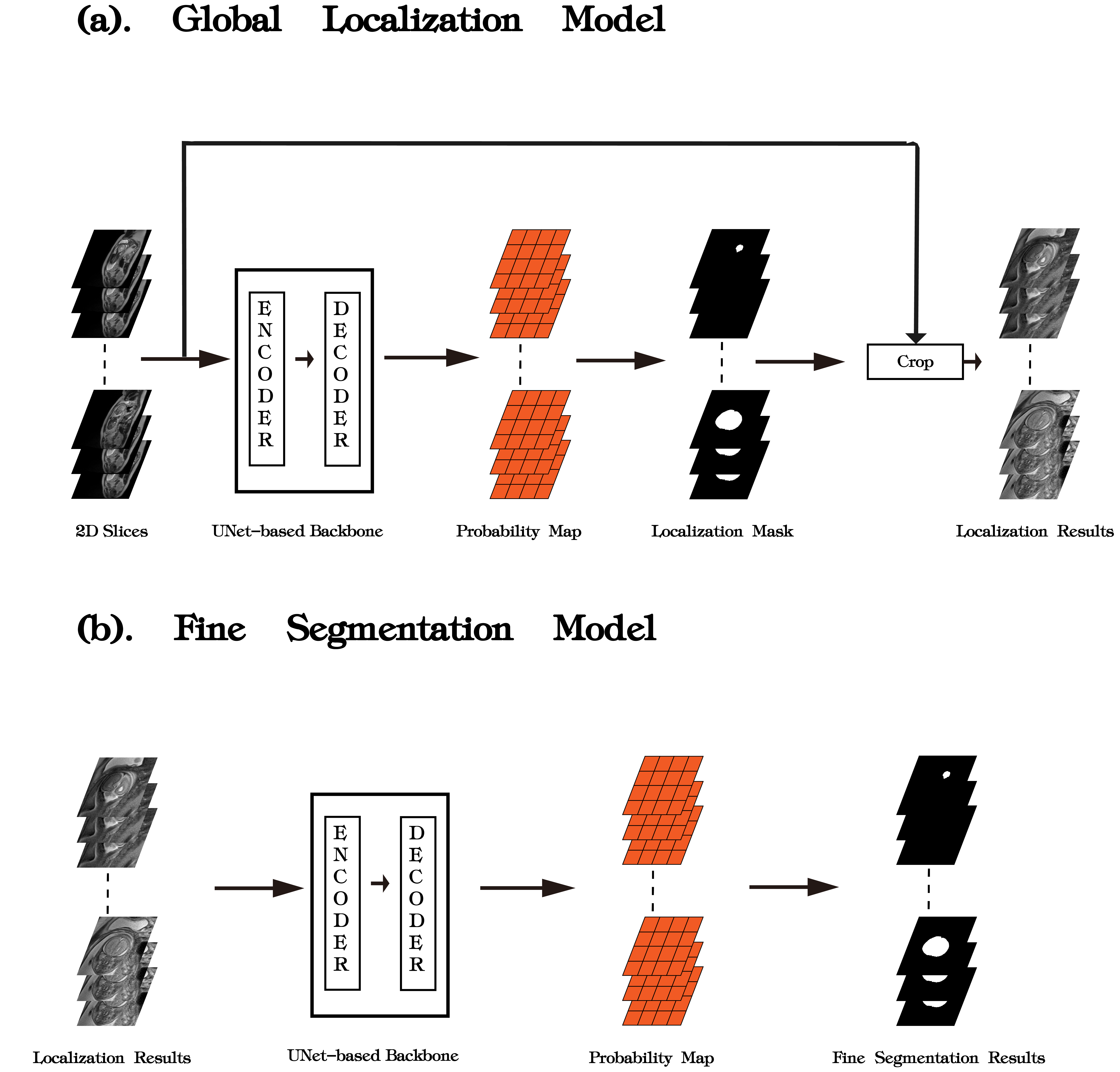}
  \caption{The cascade network structure of CasUnext}
  \label{fig:Model}
\end{figure}

The encoder pathway contains repeated application of convolution blocks to extract hierarchical features and downsample the spatial dimensions. The decoder pathway symmetrically upsamples the features and concatenates them with encoder activations using skip connections.The overall structure and the algorithmic procedure of CasUNext are illustrated in Fig~\ref{fig:CasNet} and Fig~\ref{fig:Model}, respectively. \par
The Loc-Net serves as the initial step in the segmentation process, which focuses on accurately localizing the fetal brain region within the input image. This step captures essential features and spatial information to delete redundant slices for precise localization of the fetal brain and computing resources saving. Following the Loc-Net, brain region images are generated by cropping the original image with localization labels to 256 x 256 pixels. Then the preprocessed brain region results are fed into Seg-Net, which performs a detailed analysis and segmentation of the fetal brain structures. It takes advantage of the localized brain region to extract and analyze the intricate features, enabling accurate identification and segmentation in various developmental states. During this step, the cropped brain region images also input to Loc-Net with resizing to the standard size in order to mprove the ability of generalized localization of Loc-Net.

\subsection{Loc-net/Seg-net}
\begin{figure}[h]
  \centering
  \includegraphics[width=0.45\textwidth]{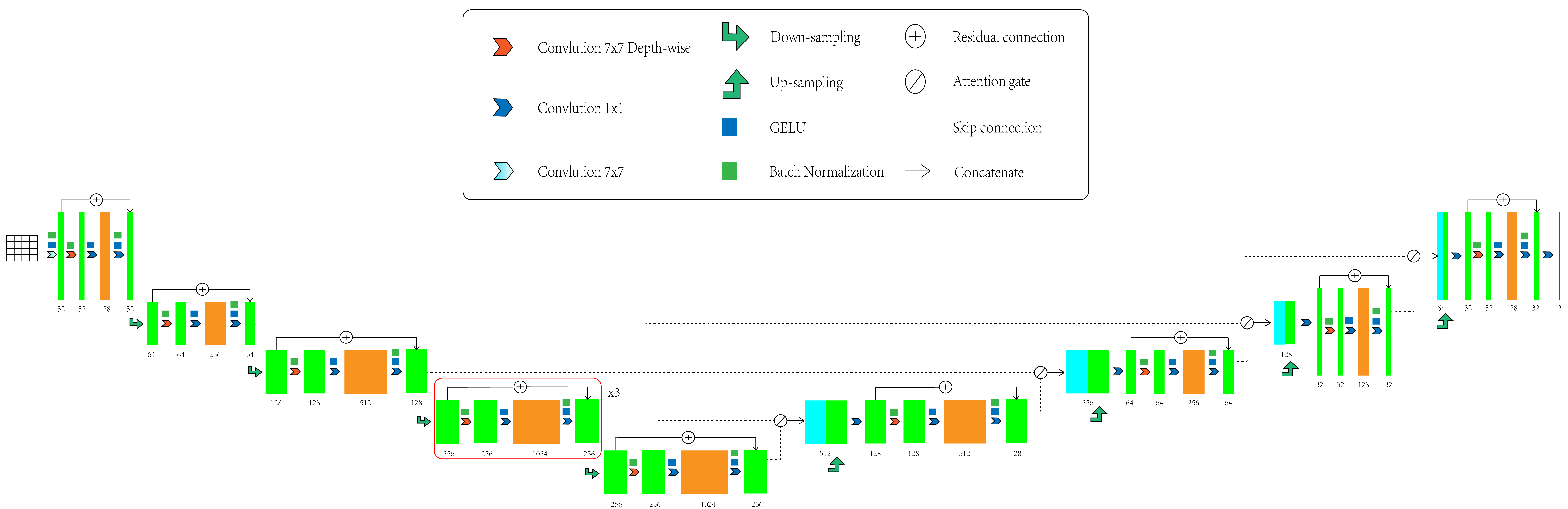}
  \caption{Loc-net/Seg-net structure}
  \label{fig:CasUNext}
\end{figure}

Both Loc-Net and Seg-Net share an overall encoder-decoder architecture based on U-Net. The detailed design of the network structure is illustrated in Fig~\ref{fig:CasUNext}.\par

\begin{figure}[h]
  \centering
  \includegraphics[width=0.45\textwidth]{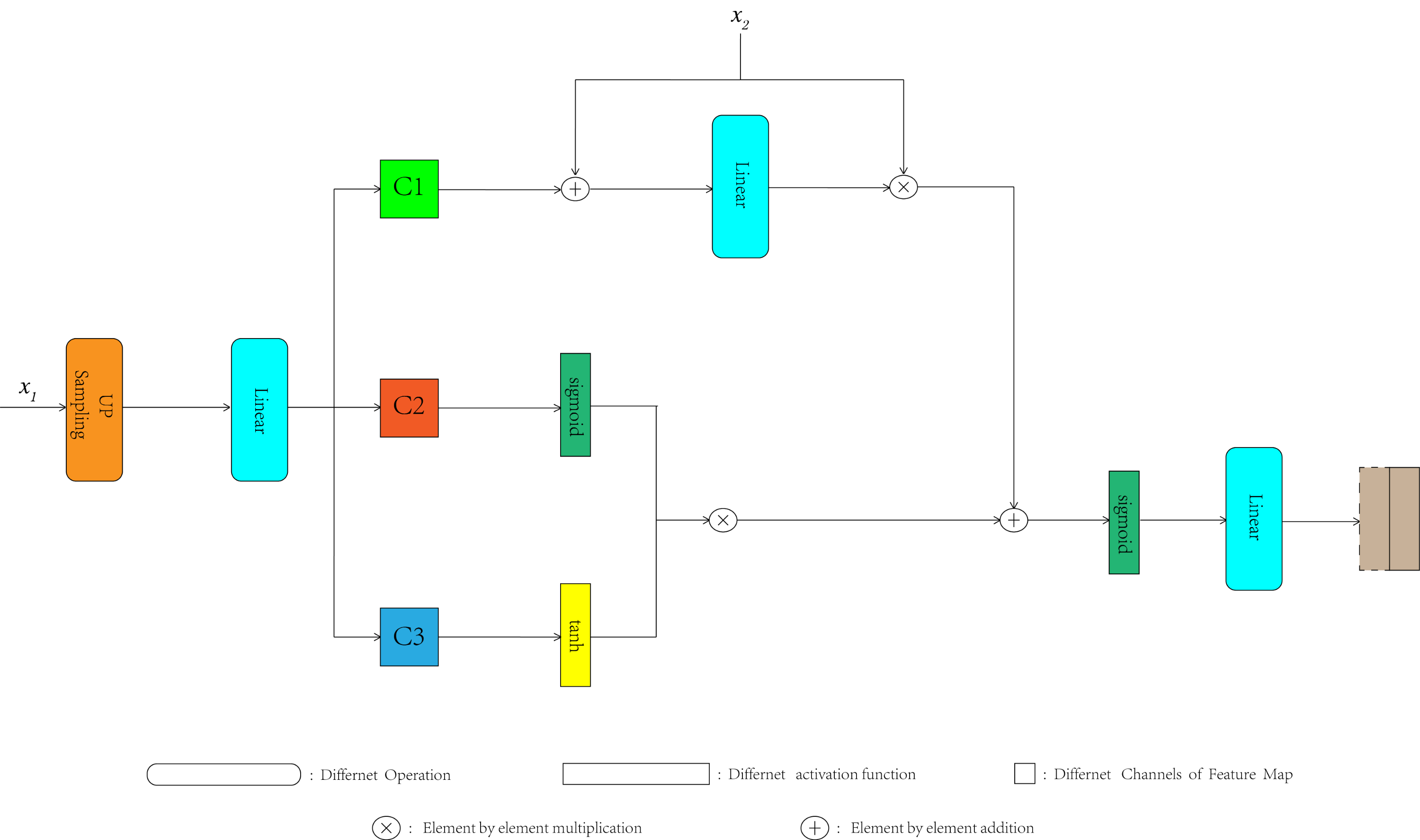}
  \caption{The structure of the Attention Gate}
  \label{fig:attention}
\end{figure}
Since the original stacks of fetal brains have a large field of view and volume generally, the input image undergoes initial processing through a convolutional layer with a kernel size of 7x7 and a stride of 1 without additional downsampling at this stage. Then the encoder contains four downsampling blocks. Each block applies 7x7 depthwise separable convolutions for feature extraction and 2x2 max pooling to halve the spatial dimensions. The number of channels are incremented as 16, 32, 64, 128 for the four blocks respectively to capture hierarchical features, and the number ratio of the last four blocks is set as 1:1:3:1 in addition. The decoder symmetrically recovers the spatial resolution using four upsampling blocks. Bilinear upsampling is followed by depthwise separable convolutions and skip connections to integrate encoder activations. Upon upsampling the feature maps, a final convolution layer is applied using point-wise convolution to map the features to the foreground and background segmentation labels.\par
In this framework, all convolutional layers, except for the initial processing step on the original input image, employ 7x7 sized convolutional kernels that leverage depthwise separable convolution. The integration of depthwise separable convolution consists of two steps: depthwise convolution and pointwise convolution, contributes to parameter reduction, computational efficiency, model size reduction, and improved accuracy.\par
At the same time, CasUNext utilize an inverted bottleneck structure, consisting of three essential steps: extended convolution, depthwise separable convolution, and linear projection. This structure enhances the efficiency of the model by reducing both the computational workload and the number of parameters in the network.\par

\subsection{Attention mechanism}
Meanwhile, we introduce a lightweight attention mechanism called Attention Gate to enhance the Skip Connection. The Attention Gate facilitates the integration of low-level semantic information during the upsampling process, enabling better capture of both low-level and high-level semantic features in medical image structures. The structure of the Attention Gate is illustrated in Fig~\ref{fig:attention}. The calculation process of the Attention Gate is:\par
\begin{gather}
  \boldsymbol{c}=upsample\left(\boldsymbol{x}_1\right)\cdot \boldsymbol{W}_1\\
  \boldsymbol{c}_1, \boldsymbol{c}_2, \boldsymbol{c}_3=split\left(\boldsymbol{c}\right)\\
  \boldsymbol{s}=\left(\boldsymbol{c}_1+\boldsymbol{x}_2\right)\cdot \boldsymbol{W}_2\\
  \boldsymbol{y}_1=\boldsymbol{s}\cdot \boldsymbol{x}_2\\
  \boldsymbol{y}_2=sigmoid\left(\boldsymbol{c}_2\right)\cdot tanh\left(\boldsymbol{c}_3\right)\\
  \boldsymbol{y}=sigmoid\left(\boldsymbol{y}_1+\boldsymbol{y}_2\right)\cdot \boldsymbol{W}_3\\
  \boldsymbol{z}=concat\left(\boldsymbol{y}, \boldsymbol{x}_1\right)
\end{gather}\par
Where $\boldsymbol{x}_1$ represents a high-level feature map, $\boldsymbol{x}_2$ represents a low-level feature map, $\cdot$ represents matrix multiplication, $upsample()$ represents up-sampling, $\boldsymbol{W}$ represents linear transformation, $split()$ represents the segmentation for feature maps in the channel dimension, $sigmoid()$$\cdot$$tanh()$ represents activation function, $concat()$ represents the splicing for feature maps in the channel dimension\par
(1) and (2) represents the high-level feature map $\boldsymbol{x}_1$ after upsampling the input linear layer, the number of channels becomes three times the original. Then it is segmented into three feature maps:$\boldsymbol{c}_1$, $\boldsymbol{c}_2$, $\boldsymbol{c}_3$.\par
(3) and (4) represents $\boldsymbol{c}_1$ makes matrix addition with the low-level feature map $\boldsymbol{x}_2$ and then passes through a linear layer. The result makes matrix multiplication with $\boldsymbol{x}_2$, which results in $\boldsymbol{y}_1$.\par
(5) represents $\boldsymbol{c}_2$ and $\boldsymbol{c}_3$ get $\boldsymbol{y}_2$ by matrix multiplication after passing through activation function.\par
(6) represents $\boldsymbol{y}_1$ makes matrix addition with $\boldsymbol{y}_2$ and then passes through the activation function and  a linear layer, the result is $\boldsymbol{y}$, which is the output of the attention mechanism module.\par
(7) represents the output of the Attention Gate $\boldsymbol{y}$ splice with $\boldsymbol{x}_1$ to get $\boldsymbol{z}$, which is the final output of the Skip Connection.\par

\section{Experiment and Results}
\subsection{Dataset and preprocessing}
The use of fetal MRI in this study was approved by the Margaret Williamson Red House Hospital Ethics Committee. The dataset comprises 150 fetal MRI scans collected from 20 gestational weeks (GWs) to 36 GWs. All MRI scans were acquired using a 1.5-T MR system by Siemens and Philips equipped with a phased-array coil. The routine MRI protocols employed for fetal assessment included axial, sagittal, and coronal T2-weighted imaging with half-Fourier single-shot turbo spin (HASTE) sequences. The imaging parameters were set as follows: TR/TE = 1350/92 ms, resolution = 0.47 × 0.47 × 4.40 mm³, and matrix size = 896 × 896 or TR/TE = 1823/92 ms, resolution = 0.55 × 0.55 × 4.40 mm³, and matrix size = 768 × 768. Manual segmentation labels were meticulously generated by experienced experts using ITK-SNAP based on the T2-weighted images serving as the ground truth for evaluating the performance of the proposed framework.\par
In this experiment, three distinct datasets were constructed to facilitate thorough evaluation. The first dataset consists of 98 coronal images from 98 pregnant women with the training set of 79 and the test set of 19. The second dataset comprises 150 pregnant women, including axial, coronal, and sagittal images with the training set of 120 and the test set of 30. Then the third dataset includes 50 abnormal fetal brain MRI all for evaluation.\par
To effectively address the data imbalance problem of different scanners, we firstly cut the edges of images to 768 × 768 if larger than that, and uniformly resized to 512 × 512 pixels, focusing on the center of the image. Furthermore, to enhance the localization capabilities of the global localization network, the centers of brains was calculated by localized labels and cropped to 256 × 256 pixels. By this processing, the dataset is standardized to a consistent size, facilitating subsequent analysis and model training.\par

\subsection{Comparison of Different Models}
To evaluate the performance of CasUNext in fetal brain segmentation, we conducted comprehensive experiments and compared its results with those of the U-Net model. The evaluation was performed on the two datasets described in this article, focusing on the localization and segmentation outcomes achieved by CasUNext and U-Net.
\subsubsection{Results on the axial, coronal, and sagittal imaging dataset}
\begin{figure}[ht]
  \centering
  \includegraphics[width=0.45\textwidth]{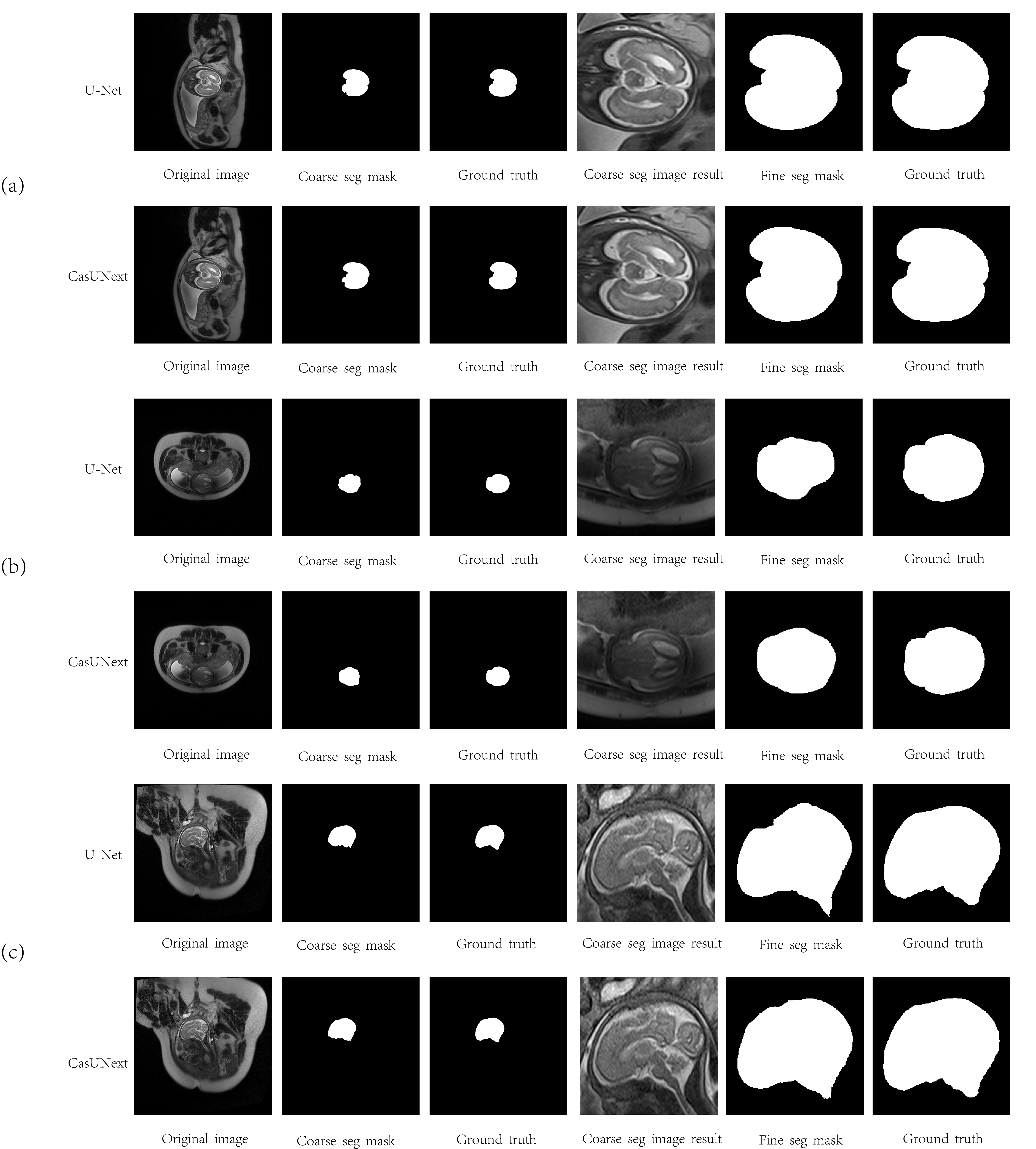}
  \caption{Comparison of results on the axial, coronal, and sagittal imaging dataset}
  \label{fig:three_combine}
\end{figure}

Results are displayed in Fig~\ref{fig:three_combine}.\par
\par
In Fig~\ref{fig:three_combine}, from left to right, the original input image, global localization prediction result (coarse segmentation result), real value of positioning step(Ground truth), coarse segmentation, fine segmentation result, and real value of segmentation step(Ground truth) are sequentially displayed. \par
Fig~\ref{fig:three_combine} (a) shows the localization and segmentation results without interference. As shown in the figure, U-Net and CasUNext can both effectively locate and segment fetal brain images without interference. \par
Fig\ref{fig:three_combine} (b) shows segmentation prediction results for images with severe motion artifacts. Under the interference of motion artifacts, The segmentation effect of U-Net has been severely affected, resulting in a large number of false negatives. CasUNext still performs image segmentation relatively accurately under the influence of motion artifacts.\par
Fig\ref{fig:three_combine} (c) shows the localization and segmentation results in the presence of interference from other body tissues of pregnant women. Under the interference of other body tissues of pregnant women, U-Net segmentation results show obvious false negatives in the upper left, and false positives in the segmentation of the brain stem in the lower right. CasUNext can accurately achieve image segmentation of fetal brain images in the presence of interference from other body tissues of pregnant women.

\subsubsection{Results on the coronal imaging dataset}
\begin{figure}[ht]
  \centering
  \includegraphics[width=0.45\textwidth]{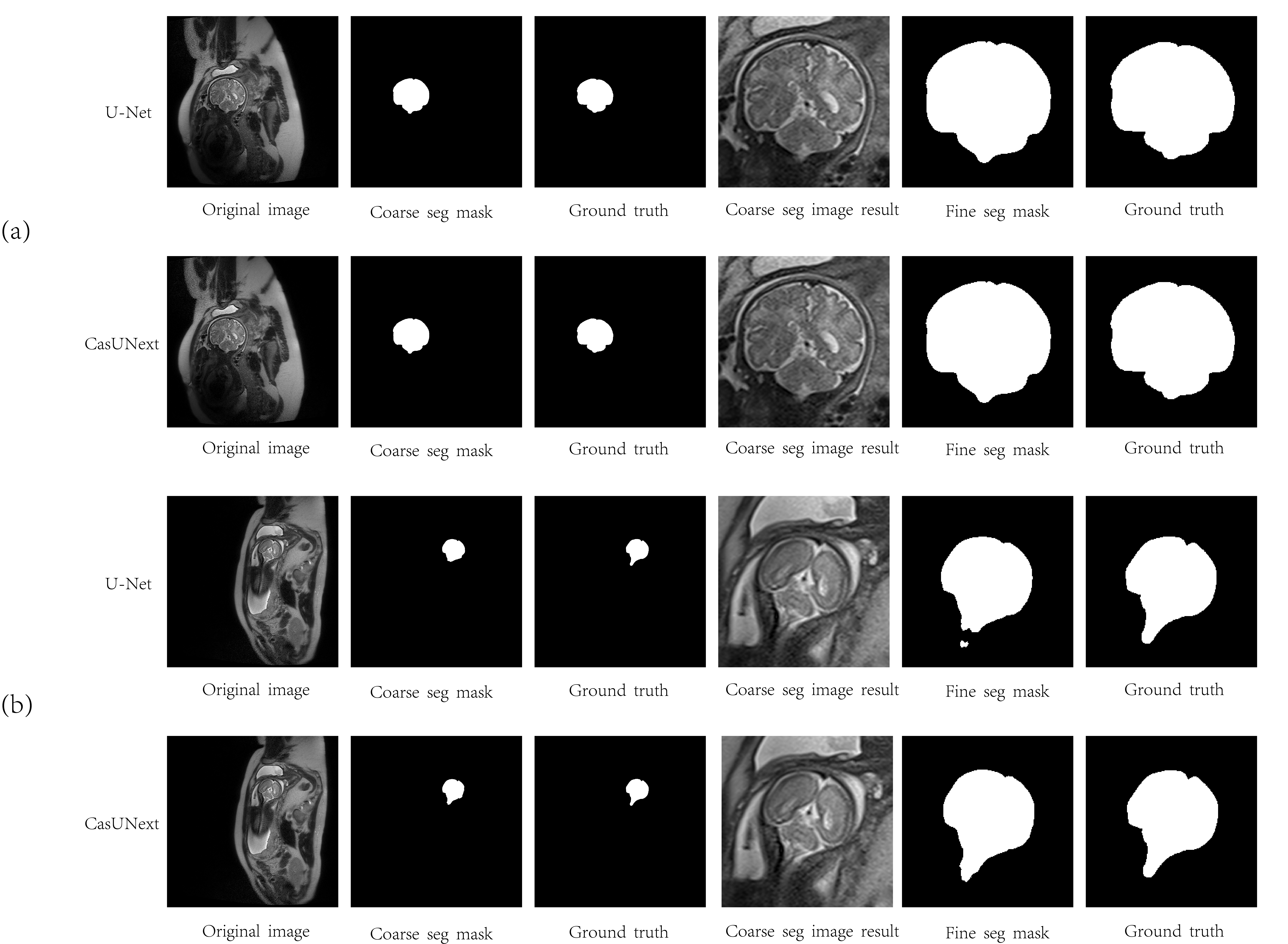}
  \caption{Comparison of results on the coronal imaging dataset}
  \label{fig:cor_combine}
\end{figure}

Results are displayed in Fig~\ref{fig:cor_combine}.\par
\par
In Fig~\ref{fig:cor_combine}, from left to right, the original input image, global localization prediction result (coarse segmentation result), real value of positioning step(Ground truth), coarse segmentation, fine segmentation result, and real value of segmentation step(Ground truth) are sequentially displayed. \par
Fig~\ref{fig:cor_combine} (a) shows the localization and segmentation results without interference. As shown in the figure, U-Net and CasUNext can both effectively locate and segment fetal brain images without interference. \par
In Fig\ref{fig:cor_combine} (b), the segmentation results for images containing brainstem. Since the fetal brain image dataset still has certain category imbalance after Data cleansing, and only a small number of 2D slices in the image sequence of each pregnant woman contain brain stem, it is easy to generate false negatives when segmenting images containing brain stem. As shown in the figure, when segmenting images containing brainstem, U-Net generated obvious false negatives, and CasUNext accurately complete image segmentation.

\subsubsection{Results on challenging condition}
\begin{figure}[ht]
  \centering
  \includegraphics[width=0.45\textwidth]{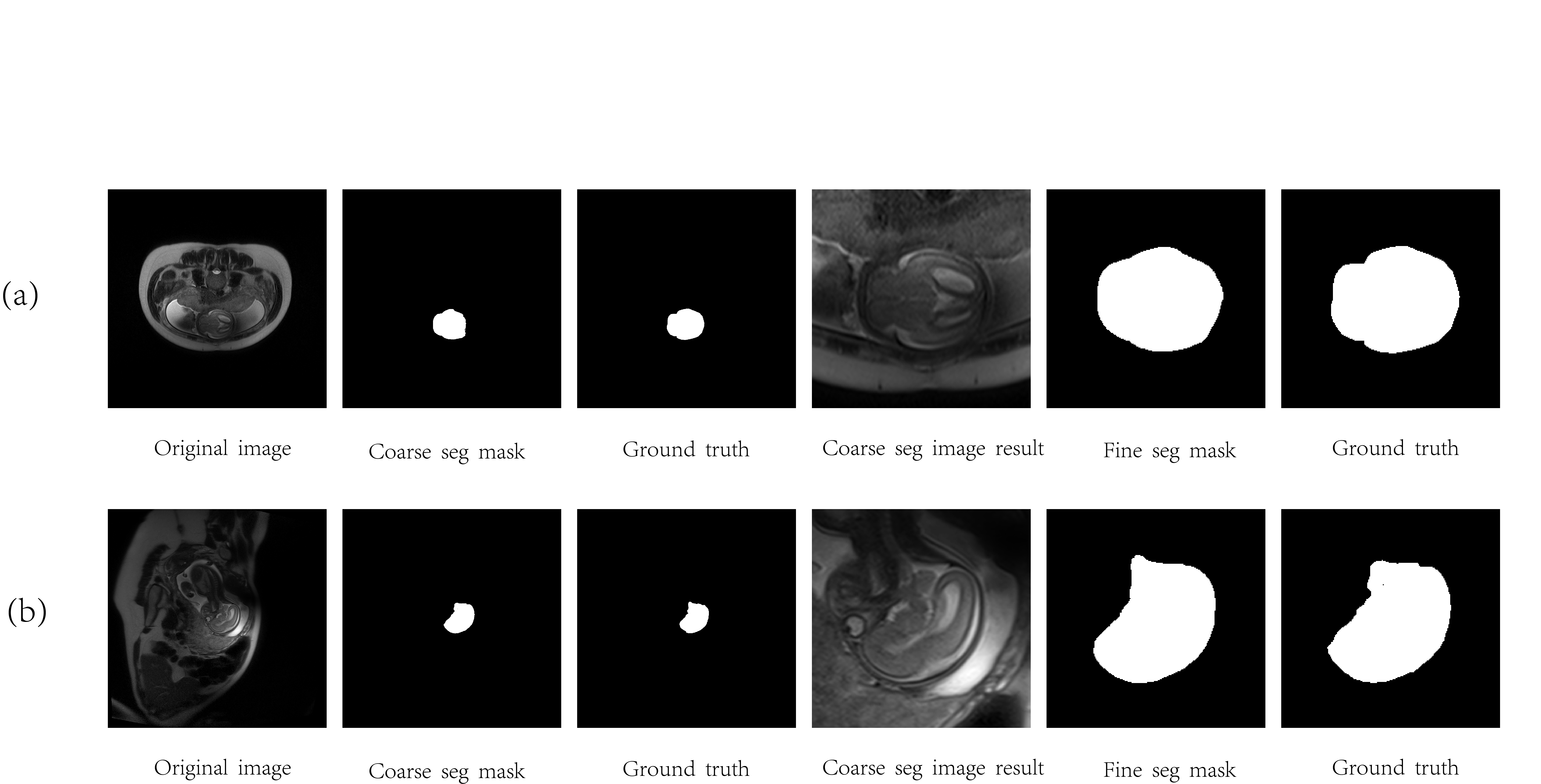}
  \caption{Comparison of results on the coronal imaging dataset}
  \label{fig:abnormal_combine}
\end{figure}

Fetal MRI are often affected by severe and arbitrary motion artifacts, and clinical datasets frequently consist of abnormal developing brains. The results on challenging condition are displayed in Fig~\ref{fig:abnormal_combine}.\par
\par
In Fig~\ref{fig:abnormal_combine}, from left to right, the original input image, global localization prediction result (coarse segmentation result), real value of positioning step(Ground truth), coarse segmentation, fine segmentation result, and real value of segmentation step(Ground truth) are sequentially displayed. \par
Fig~\ref{fig:abnormal_combine} (a) and Fig\ref{fig:abnormal_combine} (b) shows the localization and segmentation results on MRI with defects. When MRI are affected by severe and arbitrary motion artifacts or the dataset consist of abnormal developing brains, CasUNext can still accurately complete image segmentation.\par

\subsubsection{Performance Comparison}
We evaluate the performance of the model by Dice coefficient, Mean Intersection over Union (MIOU) and Sensitivity. \par
The calculation formula of Dice coefficient is shown in (8):\par
\begin{equation}
  Dice=\frac{2\ast\left|X\cap Y\right|}{\left|X\right|+\left|Y\right|} \label{eq:pingfanghe}
\end{equation}\par
Where X indicates the pixel set of the predicted image, Y indicates the pixel set of ground truth, $\left|X\right|$ indicates the number of pixels in X, $\left|Y\right|$ indicates the number of pixels in Y, $\left|X\cap Y\right|$ indicates the number of overlapping pixels between X and Y.\par
The calculation formula of MIoU is shown in (9):\par
\begin{equation}
  MIOU=\frac{1}{n}\sum_{i=1}^{n}\frac{TP_i}{TP_i+FP_i+FN_i} \label{eq:pingfanghe}
\end{equation}\par
Where n indicates the number of categories.$TP_i$ indicates true positive, $FP_i$ indicates false positive, $FN_i$ indicates false negative at the pixel level.\par
The calculation formula of Sensitivity is shown in (10):\par
\begin{equation}
  S\!ensitivity=\frac{TP}{TP+FN} \label{eq:pingfanghe}
\end{equation}\par
Where $TP$ indicates true positive, $FN$ indicates false negative at the pixel level.\par
The performance comparison between CasUNext and U-Net as global localization networks and local segmentation networks on the axial, coronal, and sagittal imaging datasets is summarized in Table 1:  \par
\begin{table}[ht]
\caption{Performance Comparison on the axial, coronal, and sagittal imaging datasets}
\resizebox{\linewidth}{!}{
\begin{tabular}{|c|c|c|c|c|c|}
\hline
Steps                                                                             & Model    & Epochs & Dice(\%)      & MIOU(\%)      & Sensitivity(\%)  \\ \hline
\multirow{4}{*}{\begin{tabular}[c]{@{}c@{}}Localization\\ (Loc-Net)\end{tabular}} & U-Net    & 100    & 84.8          & 92.6          & 88.8               \\ \cline{2-6}
& ResU-Net    & 100    & 85.4          & 92.0          & 89.3              \\ \cline{2-6}
& Attention U-Net    & 100    & 84.5          & 90.5          & 84.4               \\ \cline{2-6}
& CasUNext & 100    & \textbf{87.4} & \textbf{93.3} & \textbf{89.6}    \\ \hline
\multirow{4}{*}{\begin{tabular}[c]{@{}c@{}}Segmentation\\ (Seg-Net)\end{tabular}} & U-Net    & 300    & 91.8          & 93.2          & 93.4                        \\ \cline{2-6}
& ResU-Net    & 300    & 92.7          & 94.1          & 94.9               \\ \cline{2-6}
& Attention U-Net    & 300    & 92.4          & 93.8         & 94.3               \\ \cline{2-6}
& CasUNext & 300    & \textbf{92.9} & \textbf{94.3} & \textbf{95.6}      \\ \hline
\end{tabular}}
\end{table}

Table 1 provides a comprehensive overview of the performance metrics, including Dice coefficient, mean intersection over union (MIOU), and sensitivity, for both the localization and segmentation steps of CasUNext and U-Net. The results demonstrate that CasUNext consistently outperforms U-Net across all evaluation metrics. Specifically, CasUNext achieves a superior Dice coefficient of 87.4\% in the localization step, indicating its ability to accurately localize fetal brain regions. In the segmentation step, CasUNext achieves a remarkable Dice coefficient of 92.9\%. These results highlight the effectiveness of CasUNext in handling complex imaging scenarios. Additionally, the performance comparison on the coronal imaging dataset is presented in Table 2:  \par
\begin{table}[ht]
\caption{Performance Comparison on the coronal imaging datasets}
\resizebox{\linewidth}{!}{
\begin{tabular}{|c|c|c|c|c|c|}
\hline
Steps                                                                             & Model    & Epochs & Dice(\%)      & MIOU(\%)      & Sensitivity(\%)  \\ \hline
\multirow{4}{*}{\begin{tabular}[c]{@{}c@{}}Localization\\ (Loc-Net)\end{tabular}} & U-Net    & 100    & 90.4          & 94.1          & 91.3              \\ \cline{2-6}
& ResU-Net    & 100    & 90.9          & 94.3          & 91.0              \\ \cline{2-6}
& Attention U-Net    & 100    & 90.7          & 93.9          & 91.7             \\ \cline{2-6}
& CasUNext & 100    & \textbf{91.2} & \textbf{94.6} & \textbf{91.6}      \\ \hline
\multirow{4}{*}{\begin{tabular}[c]{@{}c@{}}Segmentation\\ (Seg-Net)\end{tabular}} & U-Net    & 300    & 95.4          & 95.3          & 94.5              \\ \cline{2-6} 
& ResU-Net    & 300    & 95.8          & 95.5          & 95.3             \\ \cline{2-6}
& Attention U-Net    & 300    & 95.5          & 95.2          & 95.0              \\ \cline{2-6}
& CasUNext & 300    & \textbf{96.1} & \textbf{95.9} & \textbf{95.9}      \\ \hline
\end{tabular}}
\end{table}
Table 2 further confirms the superior performance of CasUNext over U-Nets specifically on the coronal imaging dataset. CasUNext achieves a Dice coefficient of 91.2\% in the localization step, surpassing the performance of U-Nets and other comparative models. In the segmentation step, CasUNext achieves an outstanding Dice coefficient of 96.1\%, further establishing its capability in accurately segmenting fetal brain structures. These results indicate that CasUNext consistently outperforms U-Nets in terms of both localization and segmentation accuracy on the coronal imaging dataset. \par
The experimental results suggest that CasUNext exhibits better performance and robustness compared to U-Nets. The enhanced processing ability of CasUNext ensures accurate segmentation even in the presence of complex image characteristics. This approach showcases the effectiveness of the proposed framework in achieving precise and reliable fetal brain segmentation.

\subsection{Ablation}
To evaluate the impact of key components in CasUNext, we conducted ablation experiments by disabling specific components individually. We assessed the performance of CasUNext without the cascade network, attention mechanisms, and depthwise separable convolution. The results obtained on the axial, coronal, and sagittal imaging dataset are summarized in Table 3, while the results on the coronal imaging dataset are presented in Table 4. The performance of each ablated model is compared to that of the complete CasUNext model, revealing the contributions of the components.\par
Table 3 shows the performance comparison on the axial, coronal, and sagittal imaging dataset. CasUNext achieves a remarkable Dice coefficient of 92.9\% and a MIOU of 94.3\%. In contrast, when the attention mechanisms are removed, the model achieves a slightly lower Dice coefficient of 92.7\% and MIOU of 94.0\%. Similarly, disabling the depthwise separable convolution leads to a Dice coefficient of 92.8\% and MIOU of 93.9\%. Notably, without the cascade network, the model's performance significantly decreases, resulting in a Dice coefficient of 87.4\% and MIOU of 93.3\%. These results emphasize the significant contributions of each component in enhancing the overall performance of CasUNext.\par
Table 4 provides the performance comparison on the coronal imaging dataset. CasUNext demonstrates exceptional performance with a Dice coefficient of 96.1\% and a MIOU of 95.9\%. Conversely, removing the attention mechanisms results in a slightly lower Dice coefficient of 95.8\% and MIOU of 95.7\%. When the depthwise separable convolution is disabled, the model achieves a Dice coefficient of 96.0\% and MIOU of 95.8\%. Furthermore, without the cascade network, the model's performance significantly decreases to a Dice coefficient of 91.2\% and MIOU of 94.6\%. These findings reinforce the critical role of each component in improving the segmentation performance of CasUNext.\par
\begin{table}[ht]
\caption{Results on the axial, coronal, and sagittal imaging dataset}
\resizebox{\linewidth}{!}{
\begin{tabular}{|c|c|c|c|c|}
\hline
Dataset                                                                                       & Model             & Epochs & Dice(\%)      & MIOU(\%)      \\ \hline
\multirow{4}{*}{\begin{tabular}[c]{@{}c@{}}Axial\\ Coronal\\ Sagittal\end{tabular}} & CasUnext          & 300    & \textbf{92.9} & \textbf{94.3} \\ \cline{2-5} 
 & Without Attention & 300    & 92.7          & 94.0          \\ \cline{2-5} 
 & Without Depthwise & 300    & 92.8          & 93.9          \\ \cline{2-5} 
& Without Cascade   & 300    & 87.4          & 93.3          \\ \hline 
\end{tabular}}
\end{table}

\begin{table}[ht]
\caption{Results on the axial, coronal, and sagittal imaging dataset}
\resizebox{\linewidth}{!}{
\begin{tabular}{|c|c|c|c|c|}
\hline
Dataset                  & Model             & Epochs & Dice(\%)      & MIOU(\%)      \\ \hline
\multirow{4}{*}{Coronal} & CasUnext          & 300    & \textbf{96.1} & \textbf{95.9} \\ \cline{2-5}
 & Without Attention & 300    & 95.8          & 95.7          \\ \cline{2-5} 
 & Without Depthwise & 300    & 96.0          & 95.8          \\ \cline{2-5}  
 & Without Cascade   & 300    & 91.2          & 94.6          \\ \hline 
\end{tabular}}
\end{table}


\section{Conclusion and discussion}
In this paper, we presented a novel framework for fetal brain MRI segmentation that integrates depth-wise separable convolution, attention mechanism, and the cascade network CasUnext. Our proposed method addresses the challenges associated with fetal brain motion correction and aims to achieve high-precision brain extraction and segmentation. By leveraging the benefits of depth-wise separable convolution and attention mechanism in U-Net, we effectively integrate low-level and high-level features, resulting in improved segmentation accuracy. Through extensive experiments on a dataset of fetal brain MRI scans, we demonstrated the effectiveness of our approach. CasUnext enabled accurate segmentation results even in the presence of motion artifacts and the diverse developmental fetal brains. The quantitative evaluation, based on metrics such as dice coefficient and mean intersection over union, consistently showed superior segmentation performance compared to existing methods. The combination of depth-wise separable convolution, attention mechanism, and the cascade network yields promising results, providing accurate and efficient segmentation of fetal brains.\par
We also compare our approach to several advanced deep learning segmentation models under the same conditions, further highlighting the efficacy of our method. Our proposed framework exhibited excellent generalization properties, robustly handling multi-view imaging and abnormal fetal brain cases. The incorporation of attention mechanisms enhanced the integration of features, contributing to the overall segmentation accuracy.\par
In future work, we will expand the evaluation of our framework on larger and more diverse datasets to further validate its performance and generalizability. In addition, we will also focus on the parcellation of different brain tissues and brain regions on 2D stacks, and estimate the quantitative volumetric developmental characteristics of the brain based on precise segmentation results.\par

\bibliographystyle{IEEEtran}
\small\bibliography{refs}

\end{document}